\documentclass[10pt,twocolumn,letterpaper]{article}
\usepackage{cvpr}
\usepackage{graphicx}
\usepackage{amsmath}
\usepackage{amssymb}
\usepackage{booktabs}
\usepackage{multirow}
\usepackage[accsupp]{axessibility}

\usepackage[pagebackref,breaklinks,colorlinks]{hyperref}
\usepackage[capitalize]{cleveref}
\usepackage[nolist]{acronym}
\usepackage[numbers]{natbib}
\usepackage{collect}
\usepackage{flushend}
\usepackage{adjustbox}
\usepackage{cuted}

\crefname{section}{Sec.}{Secs.}
\Crefname{section}{Section}{Sections}
\Crefname{table}{Table}{Tables}
\crefname{table}{Tab.}{Tabs.}

\def\cvprPaperID{10420}
\def\confName{CVPR}
\def\confYear{2022}

\usepackage{soul}
\usepackage{amsmath}
\usepackage{amssymb}
\usepackage{wrapfig}
\usepackage{booktabs}
\usepackage{multirow}
\usepackage{xspace}
\usepackage{xcolor}

\def\figurePath{}

\def\myfigure#1#2{%
    \begin{figure}[tb]%
    \centering\includegraphics*[width = \linewidth]{\figurePath#1}%
    \vspace{-.3cm}%
    \caption{#2}%
    \vspace{-.3cm}%
    \label{fig:#1}%
    \end{figure}%
}

\def\mycfigure#1#2{%
    \begin{figure*}[htb]%
    \centering\includegraphics*[width = \linewidth]{\figurePath#1}%
    \vspace{-.3cm}%
    \caption{#2}%
    \vspace{-.4cm}%
    \label{fig:#1}%
    \end{figure*}%
}

\newcommand{\argmin}[1]{\underset{#1}{\operatorname{arg\,min}\ }}

\newcommand{\refSec}[1]{Sec.~\ref{sec:#1}}
\newcommand{\refFig}[1]{Fig.~\ref{fig:#1}}
\newcommand{\refEq}[1]{Eq.~\ref{eq:#1}}
\newcommand{\refTab}[1]{Tab.~\ref{tab:#1}}

\soulregister\ref7
\soulregister\cite7
\soulregister\refFig7
\soulregister\refAlg7
\soulregister\cite7
\soulregister\ref7
\soulregister\pageref7
\soulregister\shortcite7
\soulregister\eg0
\soulregister\ie0
\soulregister\etal0

\DeclareGraphicsExtensions{.png,.jpg,.pdf,.ai,.psd}
\newcommand{\mysection}[2]{\vspace{-.1cm}\section{#1}\label{sec:#2}\vspace{-.1cm}}
\newcommand{\mysubsection}[2]{\subsection{#1}\label{sec:#2}\vspace{-.1cm}}

\definecollection{mymaths}
\newcommand{\mymath}[2]{
    \newcommand{#1}{\TextOrMath{$#2$\xspace}{#2}}
    \begin{collect}{mymaths}{}{}{}{}
    #1
    \end{collect}
}

\definecolor{colorA}{HTML}{4285f4}
\definecolor{colorB}{HTML}{ea4335}
\definecolor{colorC}{HTML}{fbbc04}
\definecolor{colorD}{HTML}{34a853}
\definecolor{colorE}{HTML}{ff6d01}
\definecolor{colorF}{HTML}{46bdc6}
\definecolor{colorG}{HTML}{000000}
\definecolor{colorH}{HTML}{777777}
\definecolor{colorI}{HTML}{bdd6ff}
\definecolor{colorJ}{HTML}{6a9e6f}

\usepackage{pifont}
\newcommand{\cmark}{\checkmark}%
\newcommand{\xmark}{\scalebox{0.85}{\ding{53}}}%

\newcommand\blfootnote[1]{%
  \begingroup
  \renewcommand\thefootnote{}\footnote{#1}%
  \addtocounter{footnote}{-1}%
  \endgroup
}

\begin{document}

\begin{acronym}
\acro{STEM}{Scanning Transmission Electron Microscope}
\acro{MLE}{Maximum-likelihood Estimation}
\acro{MSE}{Mean Squared Error}
\acro{NN}{Neural Network}
\acro{NeRF}{Neural Radiance Fields}
\acro{PSF}{Point-spread Function}
\acro{PSNR}{Peak Signal-to-noise Ratio}
\acro{SIRT}{Simultaneous Iterative Reconstruction Technique}
\acro{MLP}{Multi-layer perceptron}
\end{acronym}

\mymath{\density}{\sigma}
\mymath{\position}{\mathbf x}
\mymath{\direction}{\omega}
\mymath{\ray}{\mathbf r}
\mymath{\radiance}{E}
\mymath{\pixelDistance}{p_\mathrm s}
\mymath{\circleOfConfusion}{d_\mathrm s}
\mymath{\pixel}{P}
\mymath{\area}{A}
\mymath{\directions}{\Omega}
\mymath{\blurKernel}{\kappa}
\mymath{\tiltAngle}{\alpha}
\mymath{\tiltDistance}{d}
\mymath{\residualTiltAngle}{\beta}
\mymath{\noisyRadiance}{\bar \radiance}
\mymath{\observedRadiance}{\hat \radiance}
\mymath{\noiseDensity}{p}
\mymath{\noiseGenerator}{{q}}
\mymath{\numberOfObservations}{n}
\mymath{\densityParameters}{\theta}
\mymath{\noiseParameters}{\phi}
\mymath{\loss}{\mathcal L}
\mymath{\seed}{\xi}
\mymath{\reals}{\mathbb{R}}
\mymath{\posreals}{\mathbb{R}^{+}}

\newcommand{\dataset}[1]{\textsc{#1}}
\newcommand{\method}[1]{\textcolor{color#1}{\texttt{#1}}}

\newcommand{\mypara}[1]{\noindent\textbf{#1:}\quad}

\title{Clean Implicit 3D Structure from Noisy 2D STEM Images}

\author{Hannah Kniesel\\ 
Ulm University
%{\tt\small hannah.kniesel@uni-ulm.de}
\and
Timo Ropinski\\
Ulm University
\and
Tim Bergner\\
Ulm University
\and
Kavitha Shaga Devan\\
Ulm University
\and
Clarissa Read\\
Ulm University
\and
Paul Walther\\
Ulm University
\and
Tobias Ritschel\\
University College London
\and
Pedro Hermosilla\\
Ulm University
}
\maketitle

\begin{abstract}
\acp{STEM} acquire 2D images of a 3D sample on the scale of individual cell components.
Unfortunately, these 2D images can be too noisy to be fused into a useful 3D structure and facilitating good denoisers is challenging due to the lack of clean-noisy pairs.
Additionally, representing detailed 3D structure can be difficult even for clean data when using regular 3D grids.
Addressing these two limitations, we suggest a differentiable image formation model for \ac{STEM}, allowing to learn a joint model of 2D sensor noise in \ac{STEM} together with an implicit 3D model.
%We show, that the combination of these models outperforms both individually, as well as several baselines on synthetic and real data.
We show, that the combination of these models are able to successfully disentangle 3D signal and noise without supervision and outperform at the same time several baselines on synthetic and real data.
\end{abstract}

\blfootnote{Corresponding author email address: hannah.kniesel@uni-ulm.de}

\vspace{-.3cm}
\mysection{Introduction}{Introduction}
\acp{STEM} enable the acquisition of 3D samples from  2D images on the scale of cellular components~\cite{crewe1970visibility,frank2013electron}.
This allows for addressing many important tasks in biology, that rely on the spatial organization inside cells \cite{kirkland1998advanced,pennycook2011scanning}.

In \acp{STEM}, the amount of electrons used to probe a sample needs to be low, in order to prevent sample damage as well as to keep acquisition times at bay \cite{henderson1995potential}.
This, unfortunately, leads to  2D images that can be noisy.
While many sophisticated image denoisers exist, fusing noisy 2D into consistent 3D structure poses its own challenge.
Many forms of fusing 2D information into 3D \cite{szeliski2010computer} assume that the same world point has the same properties in all its 2D projections.
In the presence of noise, this assumption does not hold.
While this might be negligible in many instances of fusing photographic-domain imagery taken under normal lighting conditions, the noise in the electron domain is much more intricate, \ie, it is, first, strong, and, second, does not follow a simple Gaussian model.
Thus, our first contribution is to model this 2D noise for \acp{STEM} using Normalizing Flows \cite{rezende2015variational,kobyzev2020normalizing} in an unsupervised setup.
\myfigure{Teaser}{Our approach enables learning of clean 3D structure from very noisy 2D sensor readings as produces by \ac{STEM}.
\vspace{-.2cm}}
To establish a link between 2D observations and a 3D model, a range of recent methods employ differentiable volume rendering \cite{tulsiani2017multi,henzler2019platonicgan}, which allow for changing 3D information so that when rendered, it matches some input. For these techniques, besides the difficulty of handling noise, representing detailed 3D structures can be challenging even for clean data, when using regular 3D grids. Fortunately, implicit models like occupancy fields \cite{chen2019learning,mescheder2019occupancy,saito2019pifu} or \ac{NeRF} \cite{mildenhall2020nerf} have recently shown great potential to represent 3D structures from photographs. These methods do not rely on a regular 3D grid, as they learn a 3D function to represents the shape itself.
The loss of this learning requires to project the 3D representation to 2D images to be compared to the 2D observations. Our second contribution is to unleash these methods for STEM reconstruction, by deriving the projection for STEM and a \ac{MLE} loss to compare the outcome to noisy observations. This does not introduce the blur arising from the L1 or L2 loss commonly used in \ac{NeRF}.

%We will apply both our contributions jointly and show that this combination can outperform both doing either individually (using \ac{MLE} without a noise model or using a noise model without \ac{MLE}) as well as several state-of-the-art baselines.
We will apply both our contributions jointly and show that this combination can successfully disentangle 3D signal and noise without supervision, and outperform at the same time both standard reconstruction algorithms and all variants of our setup where the noise is not modeled. We make all our data, code and trained networks available \footnote{https://github.com/HannahKniesel/Implicit-Electron-Tomography.git}.

\mysection{Previous Work}{PreviousWork}
\vspace{-.1cm}

\noindent \textbf{Inverse problems: }3D reconstruction from STEM images is an instance of an inverse problem.
Inverse problems aim to recover a signal from indirect measurements where the process to obtain such measurements is known.
This is modeled using a forward operator $F$ transforming the signal $x$, which we aim to recover, into the observations, $o = F(x)$.
Additionally, these observations are usually affected by noise. 

A set of well established algorithms to solve these problems are Back Projection algorithms~\cite{feldkamp1984fbp, radermacher2007weighted}.
However, such methods greatly suffer from artifacts when the number of observations is limited, which is usually the case in STEM.
Other algorithms have tried to solve the problem of limited observations using iterative algorithms~\cite{gordon1970art, andersen1984sart} with regularizers to enforce continuity on the reconstructed data such as L2, L1, or Total Variation (TV).
These problems have been studied on different fields, such as  reconstruction from electron microscope images~\cite{penczek1992three, marabini19983d, abrishami2015fast, waugh2020stemdeconv}, X-Ray computed tomography (CT)~\cite{chiffre2014ct, alder2018primaldual, baguer2020ctinv, zang2021ctnerf}, or visible light tomography~\cite{Zang_2020_CVPR}.
However, 3D reconstruction from STEM poses additional challenges such as the low number of observations, the \emph{missing wedge problem} due to the limited angle range used, and the large image sizes which translate to large memory requirements.
For a more thorough review of these 3D reconstruction methods for different electron microscope modalities we refer the reader to surveys by \citet{sorzano2017survem} and \citet{frank2013electron}.

In the last years, a new set of data-driven methods have been proposed to solve inverse problems for CT and MRI data.
These methods have addressed the problem by pre-processing the observations~\cite{anirudh2017loseviews}, post-processing the reconstruction~\cite{yang2018lowct,zhengchun2020tomogan}, learning the reconstruction process~\cite{Zhu2018mri, he2020radon}, by using iterative approaches~\cite{alder2018primaldual}, or by overfitting a neural network to a single reconstruction~\cite{baguer2020ctinv, zang2021ctnerf}.
Unfortunately, these methods have assumed a simplified noise model using a Poisson-Gaussian distribution~\cite{leuschner2021lodopab}.

%Early approaches used convolutional neural networks (CNN) to increase the number of observation~\cite{anirudh2017loseviews}, or to post-process the noisy reconstruction~\cite{yang2018lowct,zhengchun2020tomogan}.
%Other methods have learned the whole reconstruction process~\cite{Zhu2018mri, he2020radon} or learned an iterative method that gradually optimizes the reconstruction~\cite{alder2018primaldual}.
%However, these methods require large data sets for training.
%The problem of limited training data has been address by other authors where they use a neural network to overfit to a single reconstruction~\cite{baguer2020ctinv, zang2021ctnerf}.
%Unfortunatelly, these method have assumed a simplified noise model using a Poisson or Gaussian distribution~\cite{leuschner2021lodopab}.

In the field of Electron Microscopy (EM), deep learning has been recently applied to single-particle reconstruction from Cryo-EM images.
\citet{gupta2021cryogan,gupta2021cryogan2} proposed a 3D reconstruction using a volumetric representation and trained it using a GAN objective.
\citet{zhong2021cryodrgn,zhong2021cryodrgn2} recently proposed a method to represented the 3D reconstruction in Hartley space using a neural network and optimized the reconstruction and the pose information of each image together.
However, those methods rely on a large number of images covering all possible view directions for the reconstruction, and also assume a simple noise model.
Recently, a new deep learning approach has been suggested to improve reconstruction based on STEM images~\cite{han2021stemlearn}.
However, this method is composed of a denoiser network and a super-resolution module to improve the reconstructed volume obtained with standard algorithms.
In this work instead, we represent the 3D reconstruction with neural networks and learn it with a limited number of observations while simultaneously modelling the observed noise in an end-to-end framework not requiring supervision by clean images.

\noindent \textbf{Implicit reconstruction: }
Recently, representing a 3D scene implicitly using neural network has gained a lot of attention~\cite{Park_2019_CVPR, mescheder2019occnets, Michalkiewicz_2019_ICCV}.
These neural networks receive as input 3D coordinates of a point in space and output the signed distance to the surface of the object.
This concept was later extended using localized neural representations to only store information in the occupied parts of the scene~\cite{takikawa2021nglod, genova2019learning, genova2019localsdf, tretschk2020patchnets}.
Further work extended these ideas and use such representations for 3D scene reconstruction from multiple images~\cite{mildenhall2020nerf}.
They proposed a neural network to encode not only the occupancy in the scene, but also the radiance at each 3D location and output direction.
Thanks to transforming input coordinates and view direction using positional encoding, they were able to achieve high quality reconstructions.
Several works have followed up and proposed different improvements~\cite{yu2020pixelnerf, hedman2021snerg, barron2021mipnerf}.
Recent work have also used similar ideas for 3D reconstruction from different image modalities~\cite{attal2021torf}.

\noindent \textbf{Noise modelling: }
Image noise is an undesired perturbation of the measured intensity of a pixel generated by errors in the image acquisition process.
The most common noise models used in image denoising algorithms assume an additive white Gaussian noise~\cite{zhong2021cryodrgn2}, a Poisson-Gaussian model~\cite{foi2008poissongaus,zhang2019poisson,leuschner2021lodopab}, or a Gaussian distribution with pixel dependent variance~\cite{liu2014depgaus}.
Recent works have also suggested to train neural networks to denoise images in an unsupervised setting by imposing certain restrictions on the type of noise model~\cite{lehtinen2018noise2noise, batson2019noise2self, krull2019noise2void,calvarons2021imprnoise2noise,bepler2020topaz}.
Unfortunately, these simple models and strict constrains are not able to cover certain noise sources arising during the process of image generation in EM data~\cite{frank2013electron}.
In another line of research, Abdelhamed et al.~\cite{abdelhamed2019noise} have used Normalizing Flows to model the noise distribution from data in a supervised setting, without making any assumption of the underlying noise model.
In our work, we use Normalizing Flows to model the noise but we learn it in an unsupervised fashion thanks to the spatial constrains introduced by the 3D reconstruction process.

\mysection{Our Approach}{OurApproach}

\mysubsection{Overview}{Overview}
We will first present an overview of our approach (\refSec{Overview}) before describing the STEM image formation (\refSec{ImageModel}), an implicit 3D representation of the result (\refSec{GeometryModel}) and a noise model (\refSec{NoiseModel}) for STEM.
We conclude on details of the loss used (\refSec{Loss}).

\myfigure{Overview}{
    Overview of our approach.
    Given a set of $\numberOfObservations$ observations, $\observedRadiance_i$, we optimze the parameters of two tunable mappings.
    The first is an implicit 3D model (MLP, pink box) that represents the geometry as a continuous implicit density field.
    The second is a mapping from clean values of accumulate density, \ie opacity, to noisy sensor readings (Noise model, pink circle).
    3D and 2D are linked by a fixed image formation model (blue circle) and compared using a loss capable to compare distributions (MLE, green).\vspace{-.1cm}
}
Our system is modeled after \ac{NeRF} but with two important generalizations (\refFig{Overview}):
First, where \ac{NeRF} is modelling an emission-absorption model for photons \cite{mildenhall2020nerf}, we consider a model for electrons.
Second, we do not  map from a 3D solution to a single 2D image, but to a distribution of images, including the noise.
This prevents converging to the mean of the noise, which is not the correct value.
We will explain both parts in the next sections.

\mysubsection{Image formation model}{ImageModel}
Our image formation is comprised of a ray-marching variant suitable for noisy and out-of-focus electron beams.

\mycfigure{STEM}{
    \ac{STEM} imaging process:
    \textbf{(a)} Ideal double-cone set up of a single pixel.
    \textbf{(b)} Pixel distance and refocus.
    \textbf{(c)} Non-ideal double-wedge setup and circle of confusion.
    \textbf{(d)} Non-ideal tilt and resulting variation in the circle of confusion.
    \textbf{(e)} Electron transport in a slab.
    \vspace{-.4cm}
}

\mypara{Acquisition setup}
The image acquisition process in \ac{STEM} uses an electron beam that is focused at a point within the sample as it is illustrated in \refFig{STEM}, a.
Electrons that pass through the specimen from the top are then captured by the detector.
This process is repeated for all pixels in the image by displacing the sample by a certain distance, $\pixelDistance$ as seen in \refFig{STEM}, b.
Once the image is complete, the sample is tilted by \tiltAngle degrees and the spatial scan is repeated until the desired number of images is captured.
Common existing hardware allows for a tilt of up to $\pm 72$ degrees \cite{frank2013electron}.

\mypara{Raymarching for electrons}
We here adopt the pinhole emission-absorption model now often used in differentiable volume rendering \cite{henzler2019platonicgan,mildenhall2020nerf} to electron beams.

In an absorption-only model \cite{max1995optical}, the fraction of electrons lost
\begin{equation}
\frac
{\mathrm d\radiance(\ray(t))}
{\mathrm d t}
=
-\density(\ray(t))
\radiance(\ray(t))
\end{equation}
for an infinitesimal step $\mathrm d t$ at distance $t$ along a ray $\ray(\position+t\cdot\direction)$ from position \position in direction \direction is proportional to the density $\density(\ray(t))$ of the medium at that position along the ray.
Electrons, technically, are not absorbed but scattered into many different forms of secondary emissions, we ignore here as the detector is small compared to the distance to the sample, and almost all scattered electrons will not arrive at the detector. 
Also, all electrons have the same energy in \ac{STEM}, and hence density does not depend on what would be wavelength or color for photons in the optical regime.
This equation has the solution \cite{max1995optical}
\begin{equation}
\label{eq:RTESolution}
\radiance(\ray)=
\exp
\left(
-
\int_0^t
\density(\ray(t))
\mathrm d t
\right)
.
\end{equation}
The inner integral can be solved by numeric quadrature, \ie as a sum or using Monte Carlo estimation.

\mypara{Defocus}
Above considerations assume an infinitely small ray, while in reality, the contribution to the readings of a detector pixel is the confound effect of a bundle of rays.
Hence the electron beam is not a double-cone but a double-wedge as seen in \refFig{STEM}, c.
A system is in focus, if the ratio $\pixelDistance/\circleOfConfusion$ between the pixel distance and the width of the electron beam is larger than 1.
The example in \refFig{STEM}, b/c is in focus, as $\pixelDistance>\circleOfConfusion$ and so is the setup used in our results with $\pixelDistance/\circleOfConfusion=1.86$ for a tilt angle of $\tiltAngle=0$.

If the sample is tilted however (\refFig{STEM}, d), locations at distance from the tilt axis move out of the focal plane, resulting in out-of-focus blur.
While the hardware accounts for the angle of the tilt \tiltAngle, there is always a small error in angle, resulting in a residual angle \residualTiltAngle.
Not accounting for this aspect will result in a system which learns the blur or a mix of blurry and sharp observations when seeing one world point under different (residual) tilt angles.

The full out-of-focus image formation could be solved via Monte-Carlo integrating not only the path integral from \refEq{RTESolution}, but also an integral over an area of all sensor locations \area and a set of  directions \directions for a pixel \pixel
\begin{equation}
\label{eq:RTESolutionDefocused}
\radiance(\pixel) =
\int_\area
\int_\directions
\exp
\left(
-
\int_0^t
\density(\ray(\mathbf x, t))
\mathrm d t
\right)
\mathrm d \omega
\mathrm d \mathbf x
.
\end{equation}

To solve this integration problem effectively, we seek inspiration from Computer Graphics approaches for realistic simulation of lenses \cite{potmesil1981lens}, in particular screen space methods \cite{rokita1996generating}.
These represent the 5D double integral in \refEq{RTESolutionDefocused} as a 2D integral in image space instead.
This integral then becomes a spatially-varying convolution of per-pixel radiance $\radiance(\ray)$ with a \ac{PSF} of the optical system.
The action can be described by the convolution 
\begin{align}
\radiance(\pixel) &=
\blurKernel(\residualTiltAngle,\tiltDistance)
\ast
\radiance(\ray) \\
%,
%\end{equation}
%with a single blur kernel 
%\begin{equation}
\label{eq:Blurkernel}
\blurKernel
(\mathbf x)
(\tiltAngle,\tiltDistance) &= 
\exp(
-
||\mathbf x||
\cdot
\tan(\tiltAngle)
\cdot
\tiltDistance
)
\end{align}
with a single blur kernel $\blurKernel$ that depends on the tilt angle \residualTiltAngle and the image-space distance \tiltDistance from the tilt axis.
While the true \ac{PSF} of STEM might have a different shape, the qualitative low-pass is reproduced by this Gaussian which is fast to execute and well-differentiable.

\mysubsection{3D Representation}{GeometryModel}
In EM, density distribution \density in 3D space is classically modeled as a discrete grid.
We follow the recent trend \cite{mescheder2019occupancy,chen2019learning,mildenhall2020nerf} to represent such 3D fields as an implicit function instead.
We use an \ac{MLP} $\density_\densityParameters$ that maps position to density.
For details on the \ac{MLP} architecture, we refer the reader to the supplementary material.
\mypara{Notation}
Shorthand, we will drop the dependency of $\radiance(\ray)$ on \ray and will write $\radiance$ for the radiance of some ray.
On occasion, we will further write $\radiance^\densityParameters$, to denote the radiance resulting from tracing a certain ray through an implicit field defined by an \ac{MLP} chosing parameters \densityParameters.

\mysubsection{Noise model}{NoiseModel}
Unfortunately, we can only measure a noisy estimate of the true number of electrons per unit space, time and solid angle.
This is, both because of the quantized nature of the electron beam, resulting in Poisson-like noise, and due to other sources of noise, in particular from the requirement to turn the electron beam into light to be read by a photo-sensitive A/D conversion.

Instead of observing \radiance, we have to deal with samples from
$\noiseDensity(\noisyRadiance|\radiance)$, 
stating the probability density of observing \noisyRadiance when the true value is \radiance.

Were we given pairs of clean and noisy values $\noisyRadiance$ and $\radiance$ it would be simple to train a generative noise model.
In the case of \ac{STEM} however, it is difficult to acquire pairs of clean and noisy sensor readings as the sensing process itself changes the sample while at the same time depending on the sample.

As a remedy, we train the noise model jointly with the 3D density field itself.
The key insight here is that there is multiple noisy observations of the same clean density field, but under different rays.

We use Normalizing Flows \cite{rezende2015variational,kobyzev2020normalizing}, as this can both: compute the density $\noiseDensity(\noisyRadiance| \radiance)$, the probability of $\noisyRadiance$ given $\radiance$ (as required by our loss we describe next) and generate samples $\noiseGenerator(\seed|\radiance)\sim\noiseDensity$ where \seed is a random number.
We will write shorthand $\noiseGenerator^\noiseParameters$ for an instance of the noise model with tunable parameters \noiseParameters.

In particular, we use eight 1D radial flow layers~\cite{rezende2015variational} which transforms a Gaussian distribution into our desired $\noiseDensity(\noisyRadiance| \radiance)$ distribution.
These layers apply radial contraction and expansion around a reference point and are defined as:
\[
f(z) = x + \frac{\beta(z - z_0)}{\alpha + |z -z_0|}
\]
\noindent where the learned parameters are $z_0 \in \reals$, $\alpha \in \posreals$, and $\beta \in \reals$.
To condition the noise distribution to the real radiance, $\radiance$, and therefore allow for modeling signal-dependent noise, in the last four layers of our Normalizing Flow model, the learned parameters  $\{z_0, \alpha, \beta\}$ are predicted by a small \ac{MLP} which takes as input \radiance.

Recent work has suggested to use more complex Normalizing Flow models to learn a noise distribution from noise--clean pairs \cite{abdelhamed2019noise}.
Such models use CNN layers to condition the learned distribution on a region of the clean image.
Unfortunately, allowing the noise model to inspect the image could allow the model to not only learn the noise distribution but also to fix artifacts and missing details that the 3D reconstruction was not able to recover.
By conditioning the model on a single pixel, our reconstruction framework is able to separate the 3D signal from the noise.

\noindent \textbf{Implementation: }
In this section we describe the target probability density function as $\noiseDensity(\noisyRadiance| \radiance)$. 
Instead, we learn the distribution of differences between the true and the observed radiance, $\noiseDensity(\noisyRadiance - \radiance| \radiance)$.
This objective is equivalent to the one described in this section.
However, this distribution allows for the gradients to propagate not only through the flows' conditioned layers, but also directly from the loss.

\mysubsection{Loss}{Loss}
We are looking for a scalar density field $\density^\densityParameters(\position)\in\mathbb R^3\rightarrow\mathbb R$ as well as a noise model $\noiseGenerator^\noiseParameters(\seed)\in\mathbb R\rightarrow\mathbb R$ with tunable parameters \densityParameters and \noiseParameters, to explain the observed opacities $\observedRadiance_i$.
We can compute the clean solution for pixels in images as we know their relative orientations, camera geometry and hence, the ray $\ray_i$ of every pixel.

\mypara{Clean case}
With access to clean observations, we could minimize the empirical risk
\[
\argmin{\densityParameters}
\mathbb E_i
[
\loss(
\radiance_i^\densityParameters, \observedRadiance_i
)
]
.
\]
Recall, that clean samples \radiance do not exist and we have to work with noisy samples \noisyRadiance, leading to the following thought experiment:

Consider the case of a volume of constant density \densityParameters and an optimization to find this density given multiple noisy observations of that volume $\observedRadiance_i$.
If we were to minimize this under the $\loss_2$ loss, it was to produce the mean of all density solutions explaining all observation.
Under an $\loss_1$, it would converge to the median of all density solutions.
Importantly, neither mean or median of the distribution of solutions is the ground truth value \densityParameters of complex \ac{STEM} noise.

\mypara{Noisy case}
The key to make this work, is to match the entire distribution of noisy observations to a generative model producing a distribution of radiances.
The only combination to explain those distribution pairs is a parameter pair \densityParameters/\noiseParameters: a clean 3D volume that, after ray-marching and defocus blur, and after adding synthetic noise, produces the observation distribution.
Hence, we minimize \[
\mathrm{arg\,min}_{\densityParameters,\noiseParameters}
\
\mathbb E_{i,\seed}
[
\loss_\mathrm{MLE}(
\noiseDensity^\noiseParameters(\observedRadiance_i - \radiance_i^\densityParameters|\radiance_i^\densityParameters)
)
].
\]

Note, that the distribution loss is defined on the difference of noisy observations in respect to the clean ones.
This allows for learning the correct solution up to an additive constant.
As most applications are not concerned with getting the exact absolute value (such as a photo does not tell the absolute radiance unless we know exposure, aperture and ISO), this might be an acceptable limitation in most --but not all-- applications.
Noteworthy, this is not a limitation of knowing the exact physical parameters of the \ac{STEM} but a core limitation of our approach to denoising.

\mysection{Results}{Results}
We present results of our approach in different datasets, for different methods according to different metrics which we will all explain next.

\mypara{Data}
We consider a \dataset{Synthetic} and a \dataset{Real} data set.
The main motivation is, that to our knowledge no ground truth data for a quantitative evaluation of real \ac{STEM} acquisitions is available.
Recall, that training proceeds from scratch for every data set.
For every data set we have a certain number of noisy 2D images available, that is split into test, train and validation.

The \dataset{Synthetic} data set is comprised of random arrangements of ellipsoidal shells of a random density and a density model of the ZIKV (\ie, Zika) virion at 15\r{A} by \citet{long2019structural}.
As we know the clean analytic solution, this data set can be used for quantitative evaluation.
As it resembles the 3D structure of cells, it allows for qualitative evaluation, too.
We assume a 79-image tilt series ranging from -59.5$^{\circ}$ to 59$^{\circ}$ with 1.5$^{\circ}$ steps. 
Additionally, we generate 14 projections for validation, and 20 for testing. All projections are rendered in 1000$\times$1000 pixels. 
To add noise to the simulations we train a Normalizing Flow to match the noise distribution of \ac{STEM} images.
We acquire the training data of the Normalizing Flow network by real \ac{STEM} data.
During data acquisition, for each tilt step, two images are taken: One image with short exposure time and one image with long exposure time, resulting in noisy and less noisy image pairs. 
We make sure the image pairs are aligned pixel-precise, by using the SIFT algorithm of the ImageJ toolbox \cite{abramoff2004image}. To retrieve an accurate alignment, we image simple structures of nanoparticles.
Then we retrieve the difference image which is now containing only the noise of the short exposure STEM.
See \refFig{RealData} for an illustration of this process.
\myfigure{RealData}{
    To generate noisy synthetic data for evaluation, we obtain noise samples with the difference between aligned short and long exposure STEM images of a set of nanoparticles (\textbf{left}).
    We also evaluate on real STEM images of cells infected with SARS-CoV-2 where only short exposed images are available (\textbf{right}).\vspace{-.2cm}
}
We train the Normalizing Flow by conditioning it on the less noisy tilt series to then match the noise distribution of the noisy tilt series. 
By sampling from the trained Normalizing Flow model, we are able to generate a pair of tilt series, with clean and noisy projections. 

The \dataset{Real} images contain cells infected with SARS-CoV-2 using short exposure times. The tilt series ranges from -72$^{\circ}$ to 72$^{\circ}$, with a tilt step of 1.5$^{\circ}$ at a resolution of 900$\times$900 pixels. 
Before reconstruction, we align the raw tilt series with the IMOD software by \citet{kremer1996imod}, using fiducials.
For validation we use a projection of the training data. 
Without a way to capture ground truth for such data, they are used only for qualitative evaluation.

\colorlet{colorWBP}{colorA}
\colorlet{colorSIRT}{colorB}
\colorlet{colorL2Noisy}{colorC}
\colorlet{colorL2Den}{colorD}
\colorlet{colorOursSup}{colorE}
\colorlet{colorOurs}{colorF}
\colorlet{colorL2Clean}{colorG}
\colorlet{colorL2Blur}{colorH}
\colorlet{colorL1Noisy}{colorI}
\colorlet{colorWBPDen}{colorJ}

\mypara{Methods}
Besides \method{Ours}, we consider commonly used reconstruction algorithms for \ac{STEM} images and several variants of our method.
The first methods are the weighted back-projection (\method{WBP}) and the \ac{SIRT} method (\method{SIRT}) implemented in the software package IMOD \cite{kremer1996imod}, a state-of-the-art industry solution to tomographic reconstruction problems.
Next, we explore using our implicit 3D representation but training directly under and $\mathcal L_2$-loss, as done in \ac{NeRF}, for either the noisy data directly (\method{L2Noisy}) or the data denoised in 2D with a common denoiser, BM3D (\method{L2Den}).
We study a supervised variant of our approach, which assumes the knowledge of the noise model (\method{OursSup}).
Lastly, \method{L2Clean} is the same setup as \method{L2Noisy}, but trained on the clean projections instead.
This is an upper bound what the implicit 3D reconstruction can achieve for this data if no noise is present in the observations.

\newcommand{\winner}[1]{\textbf{#1}}

\begin{table*}[t]
    \centering
    \caption{Main quantitative results of different methods (rows), reconstructing a known ground truth volume according to different metrics (columns).
    The best method across all methods without access to clean ground truth are shown in \winner{bold}.}%
    \vspace{-.3cm}
    \label{tab:Results}
    \begin{minipage}[t]{12.5cm}
    \begin{tabular}{l l c c rrr rr}
        \toprule
        \multicolumn1c{Method}&
        \multicolumn1c{Loss}&
        \multicolumn1c{Imp.}&
        \multicolumn1c{Clean.}&
        \multicolumn3c{2D}&
        \multicolumn2c{3D}\\
        \cmidrule(lr){5-7}
        \cmidrule(lr){8-9}
        &
        &
        &
        &
        \multicolumn1c{\footnotesize PSNR}&
        \multicolumn1c{\footnotesize MSE}&
        \multicolumn1c{\footnotesize DSSIM}&
        \multicolumn1c{\footnotesize PSNR}&
        \multicolumn1c{\footnotesize MSE}\\
        \midrule
        \method{WBP}&
        $\mathcal L_2$&
        \xmark&
        \xmark&
        2.97&
        50.901&
        4.904&
        7.62&
        17.299
        \\
        \method{SIRT}&
        $\mathcal L_2$&
        \xmark&
        \xmark&
        3.27&
        47.622&
        4.832&
        9.13&
        12.223
        \\
        \method{L2Noisy}&
        $\mathcal L_2$&
        \cmark&
        \xmark&
        13.86&
        1.885&
        4.271&
        19.73&
        1.064
        \\
        \method{L2Den}&
        $\mathcal L_2$&
        \cmark&
        \xmark&
        18.15&
        0.849&
        1.544&
        20.25&
        0.944
        \\
        \method{Ours}&
        $\mathcal L_\mathrm{MLE}$&
        \cmark&
        \xmark&
        \winner{19.93}&
        \winner{0.645}&
        \winner{1.020}&
        \winner{21.75}&
        \winner{0.669}
        \\
        \midrule
        \method{OursSup}&
        $\mathcal L_\mathrm{MLE}$&
        \cmark&
        \xmark&
        20.07&
        0.636&
        0.991&
        20.64&
        0.864
        \\
        \method{L2Clean}&
        $\mathcal L_2$&
        \cmark&
        \cmark&
        20.73&
        0.393&
        0.840&
        21.60&
        0.691
        \\
        \bottomrule
    \end{tabular}%
    \end{minipage}%
    \begin{minipage}[t]{5cm}
    \vspace{-2cm}%
    \includegraphics[width=4.5cm]{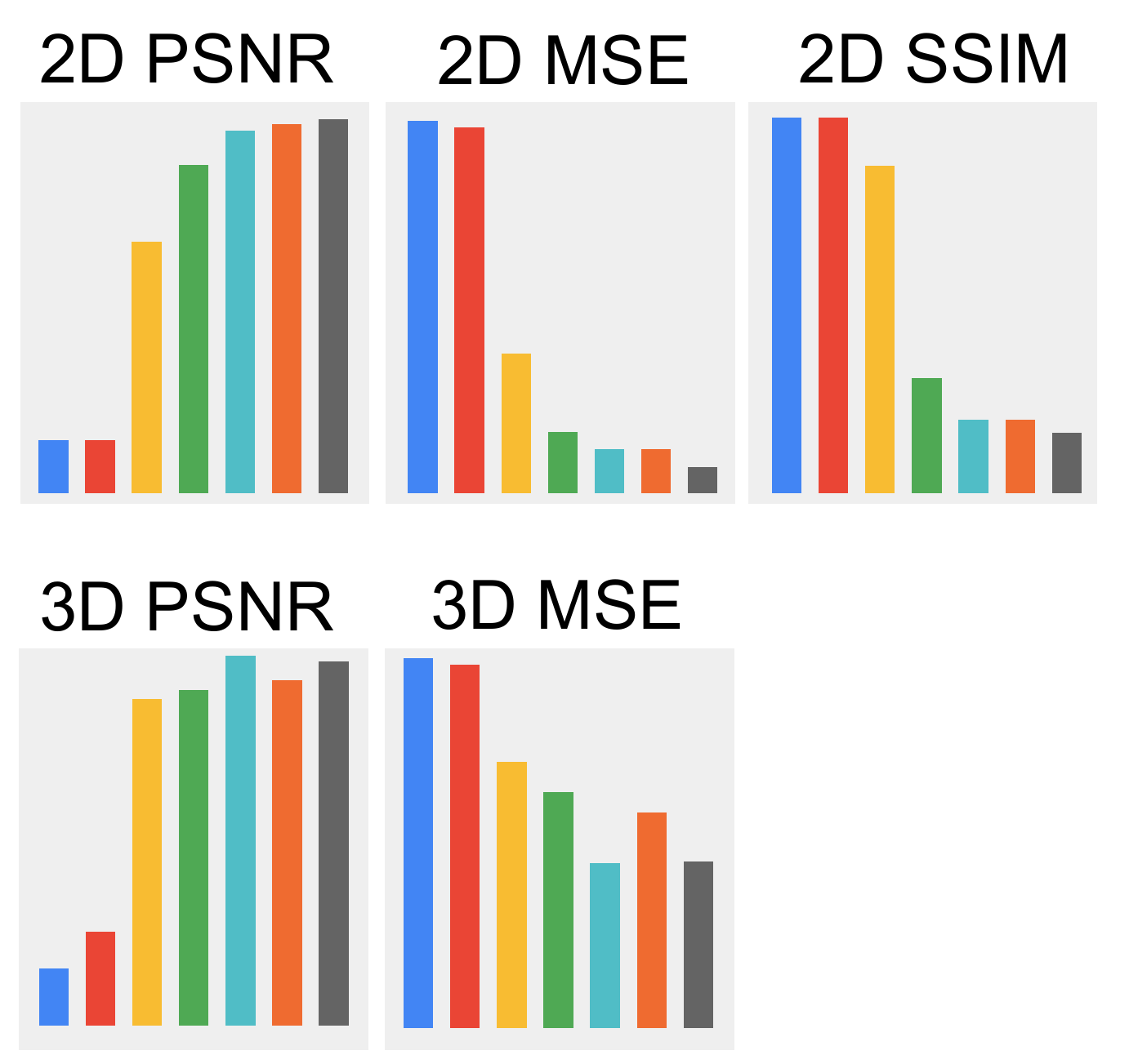}%
    \end{minipage}
    %\vspace{-.4cm}
\end{table*}%

\mycfigure{QualitativeResults_real}{
Qualitative results of different methods \textbf{(row)} for different slices \textbf{(column)} of the \dataset{Real} data set. Trends of the \dataset{Synthetic} data can be revisited on the \dataset{Real} data, even though differences in the outcome of the learned approaches are less noticeable. Again, in the presence of noise, \method{WBP} and \method{SIRT} cannot produce useful output. Training after denoising (\method{L2Den}) suppresses small details and results in an overall smoother reconstruction. But differences on \method{L2Noisy} and \method{Ours} are hard to evaluate, specially without the knowledge of GT. \vspace{-.1cm}
}

\mypara{Metric}
We consider different metrics on different forms of the results: The full 3D volume and random 2D projections.
For 2D we can apply DSSIM, \ac{PSNR} and \ac{MSE}, in 3D \ac{PSNR} and \ac{MSE}.
The full 3D volume is discretized to 1000$\times$1000$\times$1000, images are rendered in 1000$\times$1000.

\mypara{Training Details}
For training of the \ac{MLP} we use an ADAM optimizer with a learning rate of $5^{-5}$. 
For training the Normalizing Flow we found that using the SGD optimizer worked the best. 
For the \dataset{Real} data set we use 512 neurons in the hidden layers of the 3D reconstruction module, while we only use 256 neurons on the \dataset{Synthetic} data set.
For training Normalizing Flow in a supervised manner (\method{OursSup}) we used a learning rate of $5^{-7}$, whilst the training in an end-to-end fashion (\method{Ours}) required a larger learning rate of $5^{-5}$ in order to perform well. 
All networks were trained for 400,000 iterations and validation error reported every 10,000 iterations.
The model with the best validation error was chosen to compute test errors. 

\mycfigure{QualitativeResults}{
Qualitative results of different methods \textbf{(columns)} for different slices of the \dataset{Synthetic} data set \textbf{(row triplets)}.
In the presence of noise, \method{WBP} and \method{SIRT} cannot produce useful output.
Training L2 on noisy images (\method{L2Noisy}) results in blurry details (in the top triplet) as well as strong constant bias over empty space.
Training after denoising can remove this bias in empty space, but at the expense of spatial detail where the spherical virus structures have disappeared.
Our \method{OursSup} and \method{Ours} resolve  structures more similar to the ground truth.
}

\mypara{Outcome}
\refFig{QualitativeResults} presents the qualitative results of the experiments on synthetic data. 
With the Inviwo software~\cite{inviwo2019}, we created two volume renderings of the same reconstruction, one with a cut through the middle of the volume and the second one with the complete volume. Furthermore, we show a single slice of the reconstructed volume.
For the volume renderings we manually selected the transfer function on the ground truth volume to see all elements of the volume, and used it on the other methods.

We can see from the results that standard algorithms, \method{WBP} and \method{SIRT}, despite reconstructing the overall shape, miss fine details in certain areas.
The methods are not able to cope with the noise and this results in a noisy 3D reconstruction.
In \method{L2Noisy}, the noise is incorporated around the reconstructed shapes.
However, the reconstruction is over-smoothed and small details are not well recovered.
\method{L2Den} improves over \method{L2Noisy} and the noise around the volume disappears.
Unfortunately, the fine details are still not recovered.
When we look at both version of our method, \method{OursSup} and \method{Ours}, we see that the noise surrounding the volume disappears and the details are well reconstructed.
We can see that both are close to the result obtained if there is no noise present on the observations, \method{L2Clean}.
When we compare \method{OursSup} and \method{Ours} we see that \method{Ours} performs slightly better than the supervised version.

The main quantitative results are shown in \refTab{Results}.
Here we see similar results as observed in the qualitative evaluation.
\method{WBP} and \method{SIRT} obtain low performance in all metrics.
\method{L2Noisy} performs better than standard algorithms but worse than \method{L2Den}.
\method{Ours} achieves the best performance on all metrics.
When compared with the supervised version of our method, \method{OursSup}, the latest achieves slightly better performance on the 2D metrics but worse than \method{Ours} in the 3D metrics.
Lastly, the baseline \method{L2Clean} trained in the absence of noise, as expected, achieves the highest performance on almost all metrics.
However, \method{Ours} obtains a better \ac{MSE} on the 3D volume even if it is trained in the presence of noise.
This indicates that our method is not only able to model the noise, but also that the reconstruction benefits from the unsupervised setup.

Lastly, we provide the results of the qualitative evaluation on the \dataset{Real} data in \refFig{QualitativeResults_real}.
Here, we follow a similar procedure as in the \dataset{Synthetic} data set and perform a volume-based visualization and a visual analysis on the slices of the reconstructed volume.
We can see that the two baselines, \method{WBP} and \method{SIRT}, as in the synthetic data, incorporate the noise in the observation into the 3D volume, leading to reconstructions with low quality.
On the other hand, \method{Ours} is able to recover high detailed volumes while \method{L2Den} generates an over-smoothed version as when applied to synthetic data.
However, when compared to \method{L2Noisy}, even if \method{Ours} is able to better recover certain parts of the volume, the gap between these methods is smaller.
Unfortunately, the lack of a ground truth volume makes it difficult to quantitatively determine which reconstructions is more accurate.
Nevertheless, based on the qualitative evaluation and the results on synthetic data, we can conclude that \method{Ours} achieves a cleaner reconstruction.

\mypara{Ablations}
We evaluate how our framework performs under limiting model capacity for the reconstruction network.
In this experiment, we reduce the number of features in the \ac{MLP} from $256$ to $32$ and compare \method{L2Noisy} and \method{Ours} on our synthetic data set.
We can observe that when the model capacity is reduced, both methods obtain a similar reconstruction.
\method{Ours} achieved a \ac{MSE} of 1.18 on the 3D volume while \method{L2Noisy} obtained 1.07.
When the \ac{MSE} is measure on the 2D images, we obtained 4.67 for \method{Ours} and 5.40 for \method{L2Noisy}.
That might be an indication that the balance during training between 3D reconstruction and noise model requires a careful selection of the different hyper-parameters to separate 3D signal and noise.

Moreover, we evaluate the effect of accounting for the defocus in our image formation module.
We use a synthetic dataset where we add a large defocus effect based on \refEq{Blurkernel}.
We observed that accounting for this effect in the reconstruction process improves reconstruction accuracy, obtaining $0.91$ \ac{MSE} in the 3D volume instead $0.97$ \ac{MSE} when we do not account for it.
For more detailed ablation studies we refer the reader to the supplementary material.

\mypara{Limitations}%{Limitations}
Based on the ablation studies, the main limitation of our reconstruction algorithm is the careful selection of the hyper-parameters required to successfully separate 3D signal from noise.
This problem might be tackled with computational resources as is done in other reconstruction software.
However, we acknowledge that a large-scale evaluation of hyper-parameter selection on different data sets should be a future research direction.

\mysection{Conclusion}{Conclusion}
We have shown, that a combination of a noise model and an implicit 3D shape representation can acquire 3D structure from noisy observations at a quality surpassing state of the art.
To our knowledge, before no noise model for \ac{STEM} was available and no implicit representation was fit to \ac{STEM} imagery, in particular not jointly. 
Our combination makes progress along the most relevant access in this regime, the handling of noise and the representation of spatial detail.
We would hope this approach will lend itself favourable to similar high-noise, non-photographic regimes with specific noise and image formation models.

\section*{Acknowledgments}
This work was financed by the Baden-Württemberg Stiftung (BWS) for the ABEM project under grant MET-ID12–ABEM.
%"Attention-Based Segmentation and Reconstruction of Macromolecular Structures in Electron Microscopy Data"
%Renderings were produced using the Inviwo visualization framework (www.inviwo.org)

\clearpage

\setlength{\bibsep}{0.0pt}
{\small
\bibliographystyle{plainnat}
\bibliography{main}
}

\clearpage

\begin{strip}
\centering
{\Large \textbf{Supplementary material}}
\vspace{1cm}
\end{strip}

\appendix

\mysection{Model Architectures}{Supp_ModelArchitectures}
\mysubsection{Implicit Model}{Supp_ImplicitModel}

The MLP Architecture is inspired by the architecture used in NeRF by \citet{mildenhall2020nerf}. 
We forward the positional encoding of the sample's position in model space through nine fully connected layers with 256 features and a ReLU activation function.
We retrieve the output by a sigmoid activation function in the output layer to predict densities in the range $[0,1]$.
We use one skip connection, which concatenates the output of the previous layer with the input, as seen in \refFig{MLP_Architecture}.
\myfigure{MLP_Architecture}{Model architecture of the MLP used to encode the implicit reconstruction. All boxes reference fully connected layers. Light gray boxes use ReLU activation function, while dark gray boxes use sigmoid activation function. One skip connection is used by concatenating the layers output with the models input.}

\mysubsection{Noise Model}{Supp_NoiseModel}
The noise model consists of a Normalizing Flow network. This network comprises eight 1D Radial Flow layers~\cite{rezende2015variational}, of which four layers are conditioned on the clean signal. 
To condition the layers on the signal we use a MLP with one hidden layer with 16 features and ReLU activation. 
The output layer uses Tanh activation to fit the parameter range $\in [-1,1]$. 
This MLP then predicts parameters of the 1D Radial Flow layer based on the input pixel intensity.
To retrieve the noise distribution, which we want to transform into a normal distribution using the noise model, we compute the difference of the predicted clean signal and the known noisy signal.
In \method{Ours}, the clean signal is retrieved from the prediction of the implicit model since only the noisy measurement is available. 

\myfigure{NF_Architecture2}{Model architecture of the Normalizing Flow used to model the noise. Red boxes reference 1D Radial Flow layers. Green boxes reference 1D Radial Flow images, which are conditioned on the clean signal. The noise distribution is retrieved by computing the difference of the clean signal and the noisy signal. }

\mycfigure{synthetic_data}{Examples of synthetic data at low tilt angle. \textbf{Clean} is generated using our image formation model. \textbf{Noisy} adds synthetic noise to the Clean, using Normalizing Flows. \textbf{Denoised} is a denoised version of Noisy, using BM3D.}

\mysection{Synthetic Data}{Supp_SyntheticData}
To generate synthetic data we randomly place ellipsoidal shells and a density model of the ZIKV (\ie, Zika) virion at 15\r{A} by \citet{long2019structural} in a cubic volume. 
In \refFig{synthetic_data} we show an overview of the used projections for the different methods.

\mysection{Noise Synthesis}{Supp_NoiseSynthesis}
To generate the noise of the projections we train our noise model in a supervised fashion from pairs of long and short exposure STEM images, as already described in the main paper \refSec{Results}. 
We here evaluate the trained noise model in comparison to two baseline approaches: 
First, we assume a Gaussian distribution and optimize it’s parameters from the long and short exposure data using MLE.
Second, we assume a Poisson distribution and again optimize it’s parameters from the long and short exposure data using MLE.
Compared to the Gaussian, the Poisson distribution is able to model signal dependence of the noise. 
We compare the resulting noise distributions with the given distribution of the data quantitatively by reporting the Bhattacharyya coefficient and distance, as well as the Jensen-Shannon-Divergence (Table \ref{tab:noisemodel}). We also provide a qualitative evaluation in \refFig{NoiseModeling1_2}.
Both baselines seem to fit the true distribution similarly well. While the Poisson distribution prevails according to the Jensen-Shannon-Divergence, the Gaussian distribution has the overhand regarding Bhattacharyya coefficient and distance.
Still, the approximation using our noise model fits the real distribution the best in all metrics.

\mycfigure{NoiseModeling1_2}{Qualitative results of different methods \textbf{(rows)} for modelling the noise. We use MLE to approximate parameters of a Gaussian and Poisson distribution from the data. Normalizing Flow is trained on the data and makes no further assumption of the noise distribution. It outperforms the former.}

\begin{table}[h]
    \centering
    \vspace{-.2cm}
    \caption{Main quantitative results of different methods for noise modeling. We report Jensen-Shannon-Divergence (JSD), Bhattacharyya coefficient (BC) and Bhattacharyya distance ($d_\mathrm{BC}$). We optimize the parameters of the distributions Poisson and Gaussian using MLE. Our approach using Normalizing Flows outperforms the baseline methods in all metrics. The best method is shown in \textbf{bold}.}\label{tab:noisemodel}
    \begin{tabular}{l r r r}
    \toprule
    Methods & JSD & BC & $d_\mathrm{BC}$ \\\midrule
    Poisson & 0.96 & 0.99 & 0.15 \\
    Gaussian & 1.07 & 0.99 & 0.12 \\
    Normalizing Flow & \textbf{0.58} & \textbf{1.00} & \textbf{0.03}\\
    \bottomrule
    \end{tabular}
\end{table}

\mysection{Model Capacity}{Supp_ModelCapacity}
We further investigate the influence of model capacity on the performance of our method. 
Therefore, we separately investigate the capacity of the implicit model, and the noise model. 

\mysubsection{Implicit Model Capacity}{Supp_Imp_model_cap}
To investigate the effect of the capacity of the implicit model we increase and decrease the number of features in the hidden layers. 
The capacity of the noise model remains fixed during these experiments. 
We compare the performance of \method{Ours} and \method{L2Noisy}. 
We investigate performance for 32, 64, 128 and 256 features in all hidden, fully connected layers. 
We find that increasing capacity slightly improves performance of \method{L2Noisy} and \method{Ours}. Specially for small model capacities \method{Ours} is not able to outperform \method{L2Noisy} (see Table \ref{tab:MLPCap}). 

\begin{table*}[h]
    \centering
    \vspace{1cm}
    \caption{Main quantitative results of the influence of MLP Capacity. For both methods the reconstruction accuracy seems to improve with increased model capacity. Especially noteworthy is the finding, that \method{L2Noisy} outperforms \method{Ours} for small model capacities by a slight margin. We argue that this finding is mostly accountable to the imbalance of the noise model and the implicit model, since capacity of the noise model was fixed for all experiments on the MLP Capacity. Experiments on the noise model capacity underline this assumption.}\label{tab:MLPCap}
    \vspace{-.2cm}
    \begin{minipage}[t]{10.0cm}
    \begin{tabular}{l c rrr rr}
        \toprule
        \multicolumn1c{Method}&
        \multicolumn1c{Features}&
        \multicolumn3c{2D}&
        \multicolumn2c{3D}\\
        %\cmidrule(lr){5-7}
        %\cmidrule(lr){8-9}
        &
        &
        \multicolumn1c{\footnotesize PSNR}&
        \multicolumn1c{\footnotesize MSE}&
        \multicolumn1c{\footnotesize DSSIM}&
        \multicolumn1c{\footnotesize PSNR}&
        \multicolumn1c{\footnotesize MSE}\\
        \midrule
        \multirow{4}{*}{\method{L2Noisy}}&
        32&
        12.84&
        5.397&
        2.038&
        19.73&
        1.065
        \\
        &
        64&
        13.69&
        4.433&
        1.925&
        19.57&
        1.103
        \\
        &
        128&
        13.68&
        4.450&
        1.884&
        19.85&
        1.034
        \\
        &
        256&
        13.86&
        4.271&
        1.885&
        19.73&
        1.064
        \\
        \cmidrule(lr){2-7}
        \multirow{4}{*}{\method{Ours}}&
        32&
        13.47&
        4.672&
        1.990&
        19.28&
        1.181
        \\
        &
        64&
        13.32&
        4.830&
        1.925&
        19.93&
        1.017
        \\
        &
        128&
        14.15&
        3.996&
        1.912&
        19.33&
        1.166
        \\
        &
        256&
        19.93&
        1.020&
        0.645&
        21.75&
        0.669
        \\
       
        \bottomrule
    \end{tabular}%
    \end{minipage}%
    \begin{minipage}[t]{7cm}
    \vspace{-2cm}%
    \includegraphics[width=7.cm]{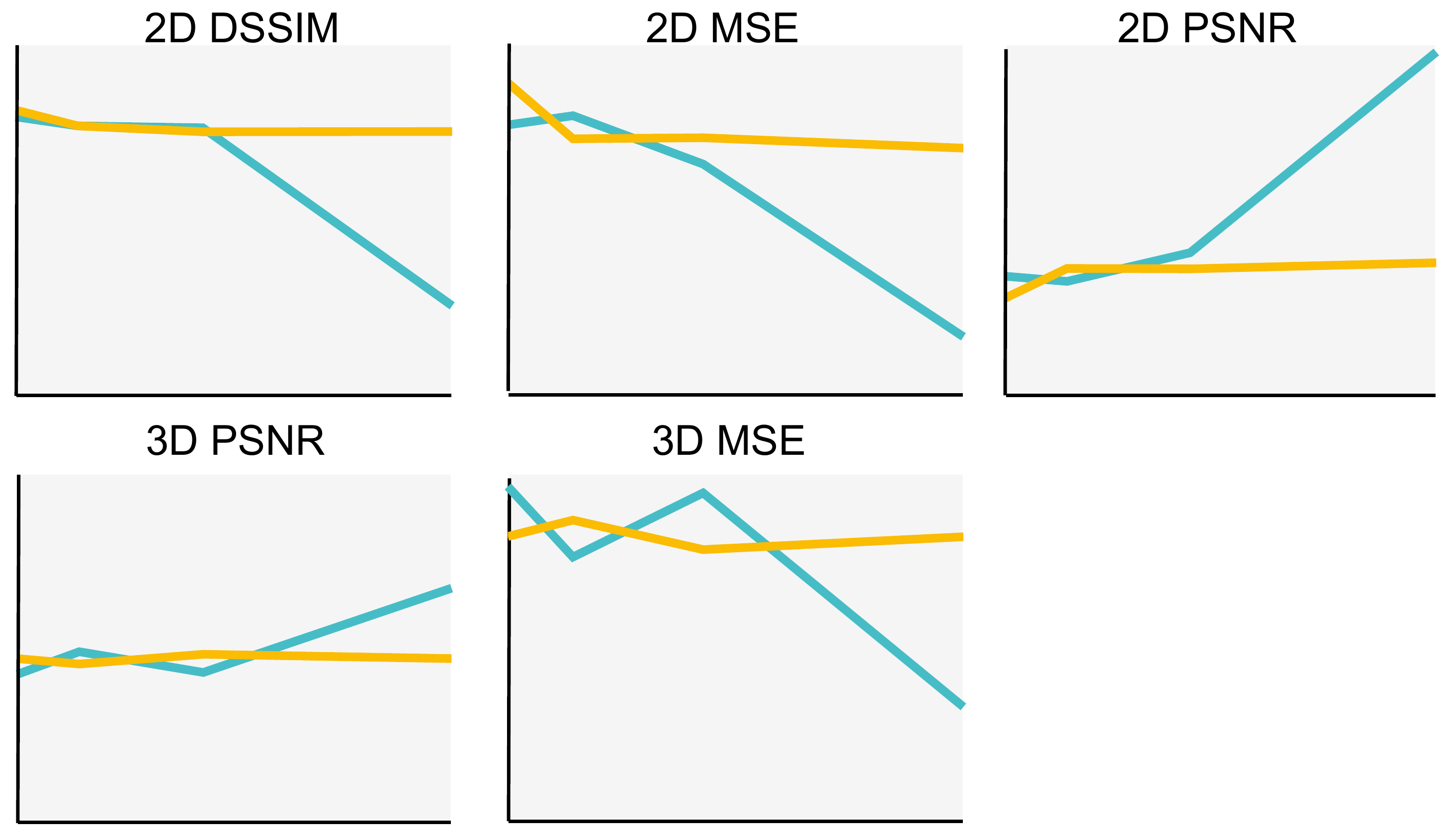}%
    \end{minipage}
\end{table*}%

\mysubsection{noise model Capacity}{Supp_noise_model_cap}
To tune the capacity of the noise model we increase and decrease the number of layers used. 
In this experiment, we reduce/increase the eight layers of the model by multiples of two. 
Still, the number of conditional and unconditional layers is always balanced. 
We fix the capacity of the implicit model to the one described in \refSec{Supp_ImplicitModel} and then train the model using \method{Ours}. 
We investigate performance of the noise model using 2, 4, 8 and 16 layers. 
Quantitative evaluation (see Table \ref{tab:NFCap}) shows that increasing the capacity of only the noise model does not necessarily improve reconstruction accuracy of \method{Ours}.

\begin{table}[h]
    \centering
    \caption{Main quantitative results of the influence of noise model Capacity. The experiment underlines the importance of balance between the noise model and the implicit model. Increasing the capacity of the noise model will not compulsorily improve performance. }
    \label{tab:NFCap}
    \begin{tabular}{c rrr rr}
        \toprule
        \multicolumn1c{Layers}&
        \multicolumn3c{2D}&
        \multicolumn2c{3D}\\
        %\cmidrule(lr){5-7}
        %\cmidrule(lr){8-9}
        &
        \multicolumn1c{\footnotesize PSNR}&
        \multicolumn1c{\footnotesize MSE}&
        \multicolumn1c{\footnotesize DSSIM}&
        \multicolumn1c{\footnotesize PSNR}&
        \multicolumn1c{\footnotesize MSE}\\
        \midrule
        2&
        13.77&
        4.352&
        1.861&
        19.77&
        1.054
        \\
        4&
        15.71&
        2.792&
        1.461&
        20.62&
        0.866
        \\
        8&
        19.93&
        1.020&
        0.645&
        21.75&
        0.669
        \\
        16&
        14.22&
        3.921&
        1.865&
        19.43&
        1.141
        \\
        \bottomrule
    \end{tabular}%
   \vspace{.1cm}
    \includegraphics[width=7.cm]{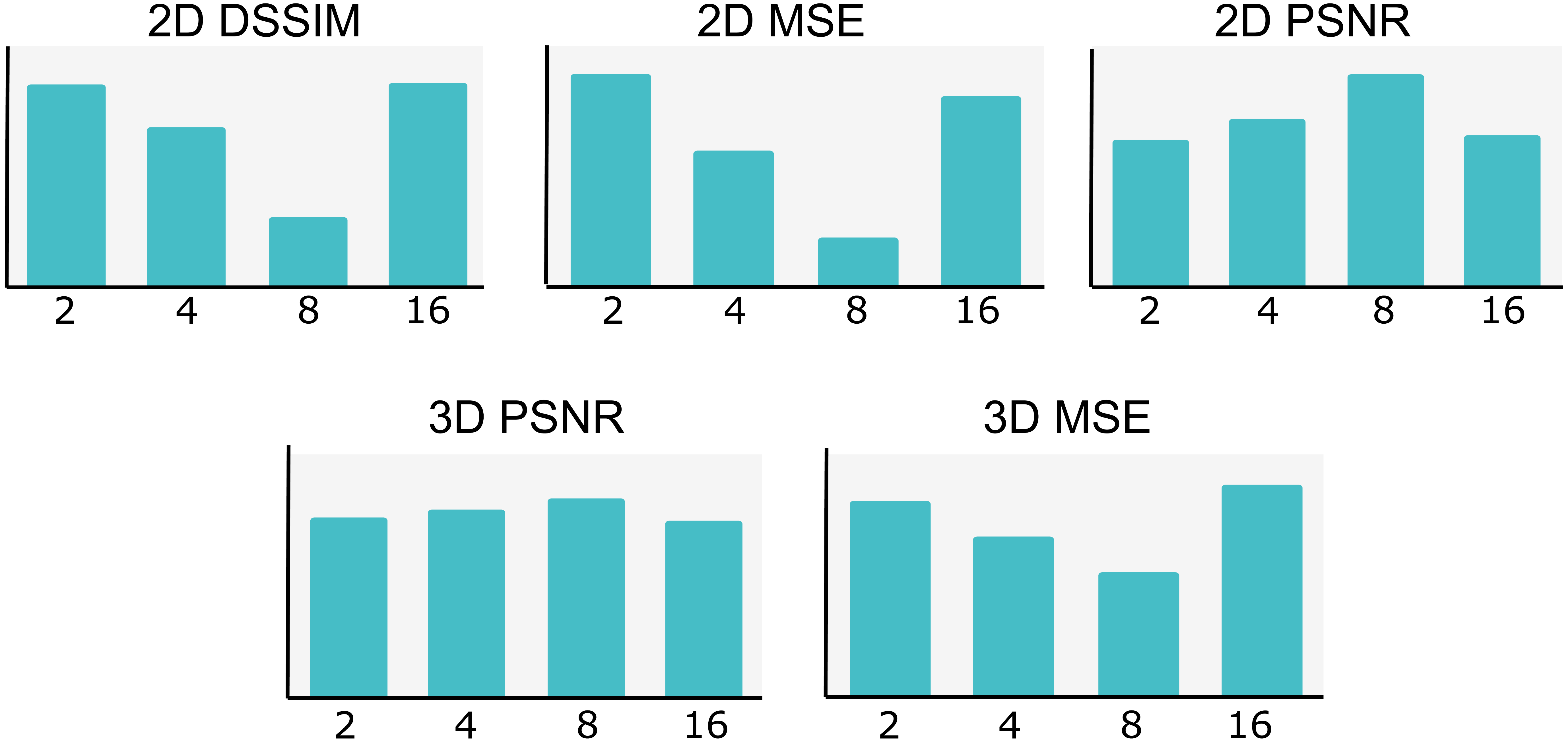}%
\end{table}%

\mysubsection{Discussion}{Supp_discussion}
\method{Ours} is able to prevent the noise model to learn structures of the 3D signal, since we only condition it on single pixels. 
On the other hand, we can not prevent the implicit model to incorporate the noise in the reconstruction by mapping it to a cylindrical structure around the region of interest as done by \method{L2Noisy}. 
Hence, we need to find a balance in training to restrict the implicit model. 
The balance between noise model and implicit model can be influenced by many different factors such as learning rate, optimizer choice, used model capacities and loss regularization terms. 
Also, approaches like alternate training, which is commonly used for the training of GAN \cite{goodfellow2014generative} models, to train Generator and Discriminator networks, might help to find a suitable balance between the noise model and the implicit model. 
We leave the investigation of these factors for future work.

\mysection{Defocus}{Supp_Blur}
During data acquisition of STEM, out-of-focus areas can occur especially at high tilt angles and at distances far away from the tilt axis. 
This can influence the reconstruction process, since observations of the same point in world space appear differently, when seen from different angles. 
We show, using synthetic data, that accounting for this blur during reconstruction can help to improve the reconstruction. 
Therefore, we apply a Gaussian blur with a variable kernel size \blurKernel, depending on the formula:
\begin{equation}
\label{eq:Supp_Blurkernel}
\blurKernel
(\mathbf x)
(\tiltAngle,\tiltDistance) = 
\exp(
-
||\mathbf x||
\cdot
\tan(\tiltAngle)
\cdot
\tiltDistance
)
\end{equation}
where $\alpha$ is the tilt angle and $d$ the distance in image space from the tilt axis. 
This formula assigns a larger kernel size to areas with high tilt angle and far distance from the tilt axis.
An example of the synthetic data in comparison to real data can be seen in \refFig{blurautocontrast}. 

During reconstruction, we apply Monte Carlo integration over the defocus area in the image by sampling multiple rays.
However, for computational reasons, we use only one sample during training.
This setup converges more slowly than using multiple samples but allows for sampling more pixels in each batch. 
We can show in a quantitative evaluation (see Table \ref{tab:blur}) that, assuming the emergence of out-of-focus blur is known, handling this blur during training improves the reconstruction quality. 

\begin{table}[h]
    \centering
    \caption{Main quantitative results of the influence of the out-of-focus effect on the reconstruction. \method{L2Clean} functions as an upper bound, as it is trained on synthetic data without out-of-focus effect. \method{L2Blur} is trained similar to \method{L2Clean} but using synthetic data which contains out-of-focus images. Lastly, \method{L2Blur}+ is trained using synthetic data with out-of-focus images, taking this into account during training. }\label{tab:blur}
    \vspace{-.2cm}
    \begin{tabular}{l l c c rrr rr}
        \toprule
        \multicolumn1c{Method}&
        \multicolumn3c{2D}&
        \multicolumn2c{3D}\\
        %\cmidrule(lr){5-7}
        %\cmidrule(lr){8-9}
        &
        \multicolumn1c{\footnotesize PSNR}&
        \multicolumn1c{\footnotesize MSE}&
        \multicolumn1c{\footnotesize DSSIM}&
        \multicolumn1c{\footnotesize PSNR}&
        \multicolumn1c{\footnotesize MSE}\\
        \midrule
        \method{L2Clean}&
        20.79&
        0.838&
        0.383&
        21.47&
        0.712
        \\
        \method{L2Blur}&
        19.66&
        1.090&
        0.514&
        20.15&
        0.965
        \\
        \method{L2Blur}+&
        20.89&
        0.819&
        0.412&
        20.42&
        0.909
        \\
       
        \bottomrule
    \end{tabular}%
\end{table}%

\mycfigure{blurautocontrast}{\textbf{Left:} Synthetic out-of-focus image at a high tilt angle. Blur is more prone for pixels further away from the tilt axis. \textbf{Right:} Out-of-focus real data for image at high tilt angle. Again, the blur is more prone in regions far away from the tilt axis.}

\mysection{Comparison of L1 and L2 Loss}{Supp_Loss}
For learned approaches which do not use a noise model, we compare the use of $L_1$ and $L_2$ loss. 
Therefore, similar to \method{L2Noisy} we train an implicit model using $L_1$ loss. We will refer to this model as \method{L1Noisy}. We found that \method{L2Noisy} outperfroms \method{L1Noisy} by a large margin. We hence used $L_2$ loss for all learned reconstructions without a noise model.  

Qualitative as well as quantitative evaluation can be seen in \refFig{Loss}. Here, we also compare to the use of the noise model by comparing to \method{Ours}.

\mycfigure{Loss}{Comparison of loss functions to compute implicit reconstruction with no noise model ($L_1$, $L_2$). \method{Ours} on the other hand uses a noise model and hence uses an \ac{MLE} loss. For reconstructions without a noise model we found that $L_2$ outperforms $L_1$. }

\mysection{Denoising of Projections}{Supp_Denoising}
We explore different denoising algorithms in order to train \method{L2Den}. 
We further evaluate the impact of denoising using WBP for reconstruction.
We compare BM3D~\cite{makinen2020collaborative}, Deep Wiener-Kolmogorov Filters~\cite{pronina2020microscopy} and Topaz Denoise (TD)~\cite{bepler2020topaz}. 
We used the provided code by the authors to apply denoising to the synthetic micrographs. For BM3D denoising we used the provided python package.
Regarding Deep Wiener-Kolmogorov Filters we assume a poisson (DWK-P) as well as a gaussian (DWK-G) noise distribution. 

\mysubsection{WBP}{Supp_WBP}
We investigate the influence of denoising the micrographs before applying WBP for reconstruction. We will call this method \method{WBPDen}. 
For results see \refFig{WBPDen}.

\mycfigure{WBPDen}{Comparison of denoisers to apply to noisy micrographs before reconstruction with WBP. We found that the reconstruction on micrographs which have been denoised with DWK-P outperforms the other denoisers regarding quantitative evaluation. Still, learned reconstructions outperform \method{WBPDen} by a large margin. }

We found that reconstruction quality was improved by all denoisers. Especially DWK-P was outperforming all other denoisers regarding quantitative evaluation. 
Still, the reconstruction quality using WBP was worse compared to all considered learned approaches. 
Moreover, we found that especially small details are not well recovered when working on denoised micrographs. 

\mysubsection{Learned Reconstruction}{Supp_learned_rec}
Similar to \refSec{Supp_WBP} we investigate the influence of denoisers on \method{L2Den}. See results in \refFig{L2Den}

\mycfigure{L2Den}{Comparison of denoisers to apply to noisy micrographs before reconstruction with \method{L2Den}. We found that the reconstruction on micrographs which have been denoised with BM3D outperforms the other denoisers regarding quantitative as well as qualitative evaluation. }

We found that BM3D outperforms all other denoisers. 
Hence, for all considered experiments of \method{L2Den} we used BM3D denoising.
Still, similar to \method{WBPDen}, small details are not well recovered when working on denoised micrographs.

\mysection{Comparison of Implicit and Explicit Reconstruction}{Supp_Type_of_Reconstruction}
We compare the benefits of using an implicit representation of the reconstruction with and without the use of a noise model. We therefore compare \method{Ours} and \method{L2Noisy} with the use of an implicit reconstruction and an explicit reconstruction accordingly. 
We initialize the explicit representation with zeros.
During training of the explicit reconstruction we also use total variation (TV) regularization. We hence compute the loss 

\begin{equation}
L = L_\mathrm{network} + \lambda \cdot L_\mathrm{TV}
\end{equation}

\mycfigure{Impl_Expl}{We evaluate the importance of using a noise model during the reconstruction, as well as the influence of using an implicit representation of the reconstruction. We evaluate this by comparing explicit and implicit reconstructions which use a noise model during training \method{Ours} and which do not use a noise model during training \method{L2Noisy}. We found that the use of an implicit representation helps to suppress artefacts generated by the missing wedge effect. Moreover, the use of a noise model seems to improve reconstruction quality.}

where $L_\mathrm{network}$ corresponds to the $\loss_2$- or \ac{MLE}- loss, depending on the used method. $L_\mathrm{TV}$ corresponds to the TV regularization which we compute on the 3D reconstruction volume. We use $\lambda$ to weight the regularization term. Based on the different scopes of the loss functions, we found that $\lambda = 0.05$ performed the best for the $\loss_2$-loss, while $\lambda = 50$ performed the best for the \ac{MLE} loss. 

For training of the explicit model without a noise model, we use an ADAM optimizer with a learning rate of $5^{-5}$. 
For the training of the explicit model using a noise model, we again use an ADAM optimizer, with a learning rate of $5^{-6}$. The noise model uses a SGD optimizer with a learning rate of $1^{-4}$. 
We again train all models for 400,000 iterations and report the test error on the model with the highest validation accuracy.

We were not able to train the explicit reconstruction on a full sized volume of shape $1000\times1000\times1000$ voxels, based on limited memory resources. We hence trained the explicit reconstruction as volume of shape $512\times512\times512$. 
During evaluation we downsample the ground truth phantom volume and we reconstruct a tomogram of similar size of the implicit model. We report PSNR and MSE in 3D on the provided tomograms. Results can be seen in \refFig{Impl_Expl}.

We found that the explicit representation is susceptible in regard of the missing wedge effect. Further, without the use of a regularization term, we observe that the explicit representation is more prone to overfit to the noise in the projections than the implicit representation. Moreover, without the use of TV regularization, the use of a noise model does not help the reconstruction. However, with the application of the TV regularization, the noise model helps the reconstruction quality. Still, the combined use of implicit representation and noise model outperforms all other baselines.

\end{document}